\documentclass[twocolumn,preprintnumbers]{revtex4-2}

\usepackage{graphicx}
\usepackage{epsf}
\usepackage{color}
\usepackage{amsmath,amssymb}
\usepackage{ascmac}
\usepackage{bm}
\usepackage{color}
\usepackage{ulem}

\newcommand{\CHSGa}{{\Sigma_g^+}}
\newcommand{\CHSGb}{{{\Sigma_g^+}'}}
\newcommand{\CHPIa}{{\Pi_u}}
\newcommand{\CHPIb}{{\Pi_u'}}
\newcommand{\CHDLa}{{\Delta_g}}
\newcommand{\CHDLb}{{\Delta_g'}}

\newcommand{\FRs}{{F_{s}(R)}}
\newcommand{\FRa}{{F_{a}(R)}}

\newcommand{\rhos}{\rho_{{\bm s}}}
\newcommand{\rhoa}{\rho_{{\bm a}}}

\newcommand{\Nadj}{{N_c^2-1}}

\usepackage{version}	% required for `\comment' (yatex added)
\begin{document}

\title{Lattice QCD study of color correlations between static quarks with gluonic excitations}

\author{Toru T. Takahashi}
\affiliation{National Institute of Technology, Gunma College, Maebashi, Gunma
371-8530, Japan}
\author{Yoshiko Kanada-En'yo}
\affiliation{Department of Physics, Kyoto University, 
Sakyo, Kyoto 606-8502, Japan}

\date{\today}

 \begin{abstract}
  We study the color correlation between
  static quark and antiquark ($q\bar q$) that is accompanied by gluonic excitations 
  in the confined phase at $T=0$ by constructing reduced density matrices $\rho$ in color space.
  We perform quenched lattice QCD calculations with the Coulomb gauge
  adopting the standard Wilson gauge action,
  and the spatial volume is $L^3 = 32^3$ at $\beta = 5.8$,
  which corresponds to the lattice spacing $a=0.14$ fm 
  and the system volume $L^3=4.5^3$ fm$^3$.
  We evaluate the color density matrix $\rho$ of static $q\bar q$ pairs
  in 6 channels ($\CHSGa$, $\CHSGb$, $\CHPIa$, $\CHPIb$, $\CHDLa$, $\CHDLb$),
  and investigate the interquark-distance dependence of color correlations.
  We find that as the interquark distance increases, 
  the color correlation quenches because of color leak 
  into the gluon field and finally approaches 
  the random color configuration in the $q\bar q$ systems 
  with and without gluonic excitations. 
  For this color screening effect, we evaluate the "screening mass" 
  to discuss its dependence on channels, the quantum number
  of the gluonic excitations.
 \end{abstract}

\maketitle

\section{Introduction}
\label{Sec.Introduction}

Color confinement is one of the most prominent features of Quantum
ChromoDynamics (QCD), and is still attracting great interest.
QCD in the confined phase gives a linearly rising potential 
between static quarks, 
and confines quarks within a totally color-singlet hadron.
Such confining features have been studied and confirmed
in a variety of approaches~\cite{Greensite:2011}.
The color confinement can be explained by a gluonic flux tube
that is nonperturbatively formed among quarks.
A color flux tube with a constant energy per length
is formed between a quark and an antiquark in the color singlet channel,
and this one-dimensional tube leads to the linearly rising
quark and antiquark ($q\bar q$) potential~\cite{Bali:1994de, Bornyakov:2004uv}.

The in-between flux tube is a colored gluonic object
created by end-point color sources, quarks.
The color charge initially associated with a {\it color-singlet}
$q \bar q$ pair flows into interquark region, and forms a flux tube 
as the interquark distance is enlarged keeping the total system color singlet
~\cite{Tiktopoulos:1976sj, Greensite:2001nx}.
This color transfer from quarks to the flux tube
can be regarded as a color charge leak from quarks to the gluon fields,
and is quantified as the color screening effect among quarks.
When interquark distance $R$ is small at $R\rightarrow 0$, color leak into gluon fields (flux tube)
hardly occurs and the quarks' color correlation is maximal.
As the interquark distance gets larger,
the gluonic flux tube is formed and quarks' color is screened by it,
which reduces the quarks' color correlation.
This correlation quench is expressed as a mixture of a ``random'' color configuration, 
in which {\it color singlet and octet components equally contribute}.
Finally at $R\rightarrow\infty$, the correlation disappears
and the quarks' color configuration is expressed solely by a random color configuration~\cite{Takahashi:2019ghj}.

In Refs.~\cite{Takahashi:2019ghj,Takahashi:2020bje}, 
the color correlation between static quark and antiquark was
investigated,
and the $R$ dependence of the color correlation for $T=0$ and $T>0$ systems was clarified in detail
by analyzing the reduced density matrix $\rho$.
In the series of papers, we also investigated entanglement entropy (EE) computed from $\rho$.
The EE under the presence of flux tube has been recently investigated in a gauge-invariant manner~\cite{Amorosso:2024leg},
and such analyses are expected to lead to further clarification of the nonperturbative aspects of QCD.

A gluonic excitation of gluon fields is also an intriguing issue
and has been studied in many situations~\cite{Juge:2002br,Muller:2019joq,Bicudo:2021tsc}.
A quark and antiquark ($q\bar q$) potential with gluonic excitations 
(hybrid $q\bar q$ potential)
was intensively studied and precisely determined, for instance, in Ref.~\cite{Juge:2002br}.
A $q\bar q$ system with  gluonic excitations
might be understood as a bound state of a color octet $q\bar q$ pair
and constituent gluons~\cite{Horn:1977rq,Ishida:1991mx,Hou:2001ig,Giacosa:2004ug,Boulanger:2008aj,Berwein:2015vca,Farina:2020slb}.
In this picture, 
quark and antiquark are expected to form a {\it color octet} configuration
at $R\rightarrow 0$, which couples with a constituent gluon
to be totally a color singlet state.
This is in contrast to the ground-state $q\bar q$ system
without gluonic excitations,
where quarks form a {\it color singlet} configuration at $R\rightarrow 0$.
Hadrons accompanied by gluonic excitations are also called hybrid hadrons,
and clarification of the internal color structure of such systems
will lead to a deeper understanding of matters obeying the strong interaction.

In this paper, we define the reduced density matrix $\rho$ for a static $q\bar q$
pair with gluonic excitations in terms of color degrees of freedom.
According to the ansatz for the reduced density matrix $\rho$
proposed in Ref.~\cite{Takahashi:2019ghj},
we investigate the color correlation inside a $q\bar q$ pair
and determine the $R$ dependence of the correlation.
In Sec.~\ref{Sec.Formalism}, we give the formalism to compute
the reduced density matrix $\rho$ of a $q\bar q$ system.
The details of numerical calculations and ansatz for $\rho$
are also shown in Sec.~\ref{Sec.Formalism}.
Results are presented in Sec.~\ref{Sec.Results}.
Sec.~\ref{Sec.Summary} is devoted to the summary and concluding remarks.

\section{Formalism}
\label{Sec.Formalism}

\subsection{reduced 2-body density matrix and $q\bar q$ correlation}

We investigate the color correlation between static quark and antiquark ($q\bar q$)
by the two-body density matrix $\rho$ evaluated in terms of $q\bar q$'s color configuration.
A density matrix $\rho$ defined in such a way is nothing but the reduced density matrix
that is obtained integrating out the other DOF ({\it e.g.} gluonic DOF)
in the full density matrix.
A reduced two-body density operator $\hat\rho(R)$ for a $q\bar q$ system
with the interquark distance $R$ is defined as
\begin{equation}
\hat\rho(R) = |\bar q(0) q(R)\rangle \langle \bar q(0) q(R)|.
\end{equation}
Here $|\bar q(0) q(R)\rangle$ represents a quantum state
in which the antiquark is located at the origin and the other quark lies at $x=R$.
The matrix elements $\rho(R)_{ij,kl}$ of the reduced density operator,
where $i$ and $k$ ($j$ and $l$) are quark's (antiquark's) color indices,
are expressed as
\begin{equation}
\rho(R)_{ij,kl} = \langle \bar q_i(0) q_j(R)|\hat\rho(R)|\bar q_k(0) q_l(R) \rangle.
\end{equation}
Then, $\rho(R)$ is an $m\times m$ square matrix with the dimension $m=N_c^2$.
The density matrix $\rho(R)$ is evaluated using only quark's color DOF, but
does not explicitly contain gluon's color DOF in this construction:
the gluonic DOF are ``integrated out'' in the lattice calculation,
and $\rho(R)$ can be regarded as a reduced density matrix.

\subsection{Ansatz for reduced density matrix $\rho_{ij,kl}(R)$}

For the time being, we consider a quark and antiquark system
{\it without} gluonic excitations.
In previous studies~\cite{Takahashi:2019ghj,Takahashi:2020bje}, we found that
the reduced color density matrix $\rho(R)$ evaluated by lattice QCD
can be precisely reproduced with an ansatz,
where $\rho(R)$ 
is expressed as a sum of an ``initial'' color singlet configuration at $R\rightarrow 0$
and mixing of the random color configuration
due to the color screening effect between quark and antiquark at finite $R$.

Let $\hat\rhos$ the density operator
for quark and antiquark forming a color singlet state
$|{\bm s}\rangle = \sum_i^{N_c} |\bar q_i q_i\rangle$
in the Coulomb gauge
as
\begin{equation}
\hat\rhos = |{\bm s}\rangle \langle {\bm s}|.
\end{equation}
In color SU($N_c$) QCD, 
the density operator $\hat\rho_{{\bm a}_i}$
for a $q\bar q$ pair in an adjoint state
$|{\bm a}_i\rangle$  is expressed as
\begin{equation}
\hat\rho_{{\bm a}_i} = |{\bm a}_i\rangle \langle {\bm a}_i|
\ \ (i =1,2,...,N_c^2-1).
\end{equation}

In the limit of $R\rightarrow 0$,
quark and antiquark are considered
to form a color-singlet state ($|{\bm s}\rangle$)
corresponding to the strong correlation limit,
since there exists no gluonic excitation in the system
and gluons form a totally colorless state.
Its density operator will be written as
\begin{equation}
 \hat\rhos
 =
 {\rm diag}(1,0,...,0)_{\alpha}.
\end{equation}
Here, the subscript ``$\alpha$'' means that the matrix is expressed
in terms of $q\bar q$'s color representation
with the vector set of $\{|{\bm s}\rangle,|{\bm a_1}\rangle,...|{\bm a_{\Nadj}}\rangle \}$.

As $R$ increases, an uncorrelated state represented by random color configurations,
where all the $N_c^2$ components mix with equal weights,
enters in $\rho(R)$ due to the color screening effect by in-between gluons.
The density operator for such a random-color state is given as 
\begin{eqnarray}
 \hat \rho^{\rm rand}
&=&
 \frac{1}{N_c^2}\left(\hat\rhos
 +
 \sum_{i=1}^{\Nadj}
\hat\rho_{{\bm a}_i}\right)
\nonumber \\
&=&
\frac{1}{N_c^2}{\hat I}
 =
\frac{1}{N_c^2}{\rm diag}(1,1,...,1)_{\alpha}.
\end{eqnarray}

Letting the fraction of the initial color singlet state being $\FRs$
and that of the random contribution being $(1-\FRs)$,
the density operator for an interquark distance of $R$ in this ansatz is written as
\begin{eqnarray}
 \hat\rho_{0,{\bm s}}^{\rm ansatz}(R)
  &=&
  \FRs\hat\rhos+(1-\FRs)\hat\rho^{\rm rand}.
\end{eqnarray}
Here, the subscript $(0,{\bm s})$ means that
the density operator $\hat\rho_{0,{\bm s}}^{\rm ansatz}(R)$ is dominated by the color singlet contribution $\hat\rhos$
at $R\rightarrow 0$.
$\hat\rho_{0,{\bm s}}^{\rm ansatz}(R)$ can be explicitly expressed as
\begin{widetext}
\begin{eqnarray}
  \hat\rho_{0,{\bm s}}^{\rm ansatz}(R)
  &=&
  \FRs\hat\rhos+(1-\FRs)\hat\rho^{\rm rand}
  \label{ansatzs01}
  \\
  &=&
  \left(\FRs+\frac{1}{N_c^2}(1-\FRs)\right) \hat\rhos
  +
  \sum_{i=1}^{\Nadj}\left(\frac{1}{N_c^2}(1-\FRs)\right) \hat\rho_{{\bm a}_i}
  \label{ansatzs02}
  \\
  &=&
  {\rm diag}\left(
	     \FRs+\frac{1}{N_c^2}(1-\FRs),\frac{1}{N_c^2}(1-\FRs)
	     ,...,
	     \frac{1}{N_c^2}(1-\FRs)
 	    \right)_{\alpha}
  \label{ansatzs03}
  \\
  &\equiv&
  {\rm diag}\left(
	     \rho(R)_{{\bm s},{\bm s}},
	     \rho(R)_{{\bm a}_1,{\bm a}_1}
	     ,...
	    \right)_{\alpha}.
\end{eqnarray}
\end{widetext}
In this ansatz,
\begin{eqnarray}
 \begin{cases}
 \rho(R)_{{\bm a}_1,{\bm a}_1}
 =
 \rho(R)_{{\bm a}_2,{\bm a}_2}
 =...=
 \rho(R)_{{\bm a}_\Nadj,{\bm a}_\Nadj}
\\  
{
 \rho(R)_{{\bm \alpha},{\bm \beta}}=0\ \ ({\rm for}\ {\bm \alpha}\neq{\bm \beta})
}
 \end{cases}
\label{Eq.conditions}
\end{eqnarray}
are satisfied at any $R$.
The normalization condition ${\rm Tr}\rho = 1$
is trivially satisfied in this ansatz as
\begin{eqnarray}
 \rho(R)_{{\bm s},{\bm s}}
+
 \rho(R)_{{\bm a},{\bm a}}
=1,
\end{eqnarray}
where 
$\rho(R)_{{\bm a},{\bm a}}\equiv \sum_{i=1}^{\Nadj}\rho(R)_{{\bm a_i},{\bm a_i}}$.
It was found that this ansatz reproduces the density matrix element 
for the ground-state $q\bar q$ system ($\CHSGa$ channel)
evaluated by lattice QCD calculation with a very good accuracy.

This ansatz can be extended to the system with gluonic excitations.
In what follows, we limit ourselves to $N_c=3$ case, hence we refer to ``adjoint'' states as ``octet'' ones.
In a $q\bar q$ system accompanied by gluonic excitations,
a $q\bar q$ pair can also form a color octet configuration 
at $R\rightarrow 0$ region,
where excited gluons carry octet color and the total system is kept color singlet.
With the density operator averaged for 
octet states
\begin{eqnarray}
\hat \rhoa &=& \frac{1}{\Nadj}\sum_{i=1}^{\Nadj}\hat\rho_{{\bm a_i}},
\end{eqnarray}
the ansatz for a $q\bar q$ system with gluonic excitations takes the form
\begin{widetext}
\begin{eqnarray}
  \hat\rho_{0,{\bm a}}^{\rm ansatz}(R)
  &=&
  \FRa\hat\rhoa+(1-\FRa)\hat\rho^{\rm rand}
  \label{ansatza01}
  \\ 
 &=&
  \left(\frac{1}{N_c^2}(1-\FRa)\right) \hat\rho_{{\bm s}}
  +
  \sum_{i=1}^{\Nadj}\left(\frac{1}{\Nadj}\FRa+\frac{1}{N_c^2}(1-\FRa)\right) \hat\rho_{{\bm a}_i}
  \label{ansatza02}
  \\ 
 &=&
  {\rm diag}\left(
	     \frac{1}{N_c^2}(1-\FRa)
             ,\frac{1}{\Nadj}\FRa+\frac{1}{N_c^2}(1-\FRa)
	     ,...
 	    \right)_{\alpha}
  \label{ansatza03}
\end{eqnarray}
\end{widetext}
Here, $\FRa$ represents the fraction of the initial color octet state,
and the subscript $(0,{\bm a})$ means that
the density operator $\hat\rho_{0,{\bm a}}^{\rm ansatz}(R)$ is dominated by the color octet contribution $\hat\rhoa$
at $R\rightarrow 0$.

In the later sections, we see that this ansatz actually
reproduces the color density matrix for a static $q\bar q$ system 
with gluonic excitations.

\subsection{Lattice QCD formalism}

In a static $q\bar q$ system, there exist three quantum numbers.
One is the eigenvalue of the projected total angular momentum
$\Lambda\equiv {\bm J}\cdot \hat{\bm r}$,
where ${\bm J}$ is the total angular momentum gluon fields possess
and ${\bm r}$ is the unit vector parallel to the $q\bar q$ axis.
We assign capital Greek letters $\Sigma, \Pi, \Delta$,..
for $\Lambda = 0,1,2,..$ states.
Second is the eigenvalue $\eta_{CP}$
of the simultaneous operations of charge conjugation 
and spatial inversion about the midpoint between $q\bar q$.
States with $\eta_{CP}=1(-1)$ are denoted by the subscripts $g(u)$.
Third is a label adopted only for the $\Sigma$ channel.
Even (odd) $\Sigma$ states
under the reflection in a plane that contains the $q\bar q$ axis
are represented by the superscripts $+(-)$.
The first excited state in each channel is discriminated by a prime mark.
Then, the quantum numbers for each state investigated in this paper are denoted as $\Gamma = \Lambda_{\eta_{CP}}^{\pm(')}$

The position on the lattice site is denoted as
${\bm r}=(x,y,z,t)=x{\bm e_x}+y{\bm e_y}+z{\bm e_z}+t{\bm e_t}$,
and the $\mu$-direction ($\mu=x,y,z,t$) link variable at ${\bm r}$
is expressed as $U_\mu({\bm r})$.
With a lower staple $S^L(R,T;\Gamma)$ representing
$q\bar q$ pair creation and propagation and 
an upper staple $S^U(R,T;\Gamma)$ for $q\bar q$ pair annihilation
that are defined as
\begin{eqnarray}
 S^{(n)L}_{ij}(R,T;\Gamma)\equiv
\left(
\prod_{t=-1}^{-T/2} U_{t}^\dagger(t{\bm e_t})
{\cal O}^{(n)}_\Gamma\left(R,-\frac{T}{2}{\bm e_t}\right)\right. \nonumber
\\ 
\left. \times \prod_{t=-T}^{-1} U_{t}(R{\bm e_x}+t{\bm e_t})
\right)_{ij},
\end{eqnarray}
\begin{eqnarray}
 S^{(n)U}_{ij}(R,T;\Gamma)\equiv
\left(
\prod_{t=0}^{T/2-1} U_{t}(t{\bm e_t})
{\cal O}^{(n)}_\Gamma\left(R,+\frac{T}{2}{\bm e_t}\right)\right. \nonumber
\\
\left. \times
\prod_{t=T-1}^{0} U_{t}^\dagger(R{\bm e_x}+t{\bm e_t})
\right)_{ij},
\end{eqnarray}
we compute $L_{ij,kl}^{(m,n)}(R,T;\Gamma)$ as
\begin{equation}
 L^{(m,n)}_{ij,kl}(R,T;\Gamma)\equiv
  S^{(m)U}_{ij}(R,T;\Gamma)S^{(n)L\dagger}_{kl}(R,T;\Gamma).
\end{equation}
Here, $q\bar q$-state creation/annihilation operator
${\cal O}^{(n)}_\Gamma(R,T)$ that connects $x=0$ and $x=R$ sites at $t=T$
is constructed so that it creates a $q\bar q$ state
with a quantum number $\Gamma$~\cite{Juge:2002br,Muller:2019joq}.
For example, when $\Gamma=\CHSGa$, we simply adopt
\begin{eqnarray}
{\cal O}^{(n)}_{\CHSGa}(R,T)
\equiv
\prod_{x=0}^{R-1}  U^{(n)}_{x}(x{\bm e_x}+T{\bm e_t}).
\end{eqnarray}
The superscripts $(n)$ denote the smearing levels
for spatial link variables.
We finally construct the operator $L_{ij,kl}^{[n]}(R,T;\Gamma)$
which encodes the elements of the density matrix $\rho$ 
for $n$-th excited state in the $\Gamma$ channel as
\begin{eqnarray}
L_{ij,kl}^{[n]}(R,T;\Gamma)\equiv
\sum_{m',n'}C^{[n]}_{m'}C^{[n]}_{n'} L^{(m',n')}_{ij,kl}(R,T;\Gamma),
\end{eqnarray}
where the coefficients $C^{[n]}_{m}$ are determined so as to maximize
the overlap of the creation/annihilation operator
$\sum_mC^{[n]}_{m}{\cal O}^{(m)}_{\Gamma}(R,T)$
to $n$-th excited state signals.
The operator optimization is done by solving a generalized eigenvalue problem~\cite{Luscher:1990ck,Perantonis:1990dy,Bicudo:2021tsc}
for $q\bar q$ potentials in the target channel $\Gamma$.
When $L_{ij,kl}^{[n]}(R,T;\Gamma)$ couples only to the $n$-th excited state
in the $\Gamma$ channel,
$\langle L^{[n]}_{ij,kl}(R,T;\Gamma) \rangle$ is then expressed as
\begin{eqnarray}
&&
\langle L^{[n]}_{ij,kl}(R,T;\Gamma) \rangle
\nonumber \\
&=&
C
\langle q(0)\bar q(R)|
e^{- \frac12\hat{H} T} |q_i(0) \bar q_j(R)\rangle
\nonumber \\
&\times&
\langle \bar q_k(0) q_l(R)|e^{- \frac12\hat{H} T}
| q(0)\bar q(R)\rangle
\nonumber \\
&=&
Ce^{- E_n T}\langle q(0)\bar q(R)|q_i(0) \bar q_j(R)\rangle\langle \bar
q_k(0) q_l(R)| q(0)\bar q(R)\rangle
\nonumber \\
&=&
Ce^{- E_n T}\rho(R)_{ij,kl},
\end{eqnarray}
where $E_n$ is the $n$-th excited-state energy.
Normalizing $\langle L^{[n]}(R,T;\Gamma) \rangle$ so that
${\rm Tr}\ \langle L^{[n]}(R,T;\Gamma) \rangle =\sum_{ij} \langle L^{[n]}_{ij,ij}(R,T;\Gamma)
\rangle = 1$,
we finally obtain the two-body color density matrix $\rho(R)$ 
for the $n$-th excited state in the $\Gamma$ channel
whose trace is unity (${\rm Tr}\ \rho(R)=1$).

\subsection{Lattice QCD parameters}

We performed quenched calculations for 
reduced density matrices of static quark and antiquark ($q\bar q$) systems
adopting the standard Wilson gauge action.
The gauge configurations are generated
on the spatial volume of $V = 32^3$ with the gauge couplings $\beta = 5.8$,
which corresponds to $V = 4.5^3$ [fm$^3$]. The temporal extent is $N_t = 32$.
All the gauge configurations are gauge-fixed with the Coulomb gauge condition.
While finite volume effects would still remains for $V = 4.5^3$ [fm$^3$]~\cite{Takahashi:2019ghj}, 
detailed study taking care of such systematic errors 
is beyond the scope of the present paper, 
in which the color structure of a $q\bar q$ pair with gluonic excitations
is clarified for the first time.

\section{Lattice QCD results}
\label{Sec.Results}

\subsection{Energy spectrum}

\begin{figure}[h]
\includegraphics[width=08cm]{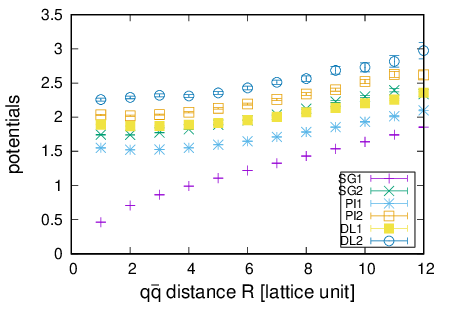}
\caption{\label{spectra}
The energy spectra for $\Gamma=$ $\CHSGa$[SG1], $\CHSGb$[SG2], $\CHPIa$[PI1], $\CHPIb$[PI2], $\CHDLa$[DL1] and $\CHDLb$[(DL2]
plotted as a function of interquark distance $R$ in lattice unit.}
\end{figure}

In Fig.~\ref{spectra},
the energy spectra for $\Gamma=\CHSGa$, $\CHSGb$, $\CHPIa$, $\CHPIb$, $\CHDLa$ and $\CHDLb$
are plotted as a function of interquark distance $R$.
The state with the lowest excitation is the $\CHPIa$ state,
and $\CHDLa$ and $\CHSGb$ states are the second-lowest excited states
that almost degenerate in energy.
The $\CHPIb$ state's energy is higher than theirs.
These spectra can be compared with the data of the preceding lattice QCD calculation~\cite{Juge:2002br} and the data obtained in this study are found to be consistent with them within errors. 
The energy for $\CHDLb$ channel,
which was not computed in Ref.~\cite{Juge:2002br},
lies above that for $\CHPIb$ channel.

\subsection{Singlet and octet components}

\begin{figure}[h]
 \includegraphics[width=8cm]{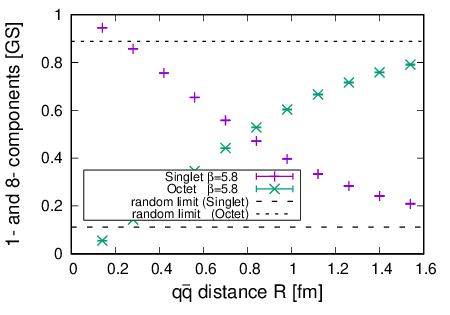}
\caption{\label{SO_SG1}
Singlet and octet components ($\rho_{\bm s, \bm s}$ and $\rho_{\bm a, \bm a}$) for $\Gamma=\CHSGa$ channel plotted as a function of the interquark distance $R$. The dashed and dotted lines indicate the values 1/9 and 8/9 for $\rho_{\bm s, \bm s}$ and $\rho_{\bm a, \bm a}$ in the random-limit color configuration, respectively.}
\end{figure}

\begin{figure}[h]
 \includegraphics[width=8cm]{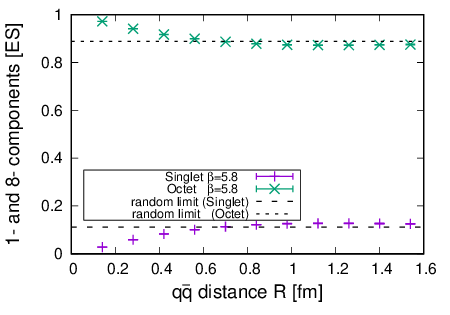}
\caption{\label{SO_SG2}
Singlet and octet components ($\rho_{\bm s, \bm s}$ and $\rho_{\bm a, \bm a}$) for $\Gamma=\CHSGb$ channel plotted as a function of the interquark distance $R$. The dashed and dotted lines indicate the values 1/9 and 8/9 for $\rho_{\bm s, \bm s}$ and $\rho_{\bm a, \bm a}$ in the random-limit color configuration, respectively.}
\end{figure}

\begin{figure}[h]
 \includegraphics[width=8cm]{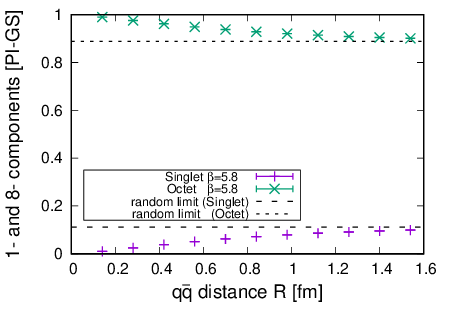}
\caption{\label{SO_PI1}
Singlet and octet components ($\rho_{\bm s, \bm s}$ and $\rho_{\bm a, \bm a}$) for $\Gamma=\CHPIa$ channel plotted as a function of the interquark distance $R$. The dashed and dotted lines indicate the values 1/9 and 8/9 for $\rho_{\bm s, \bm s}$ and $\rho_{\bm a, \bm a}$ in the random-limit color configuration, respectively.}
\end{figure}

\begin{figure}[h]
 \includegraphics[width=8cm]{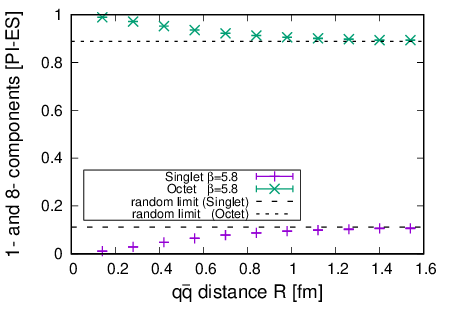}
\caption{\label{SO_PI2}
Singlet and octet components ($\rho_{\bm s, \bm s}$ and $\rho_{\bm a, \bm a}$) for $\Gamma=\CHPIb$ channel plotted as a function of the interquark distance $R$. The dashed and dotted lines indicate the values 1/9 and 8/9 for $\rho_{\bm s, \bm s}$ and $\rho_{\bm a, \bm a}$ in the random-limit color configuration, respectively.}
\end{figure}

\begin{figure}[h]
 \includegraphics[width=8cm]{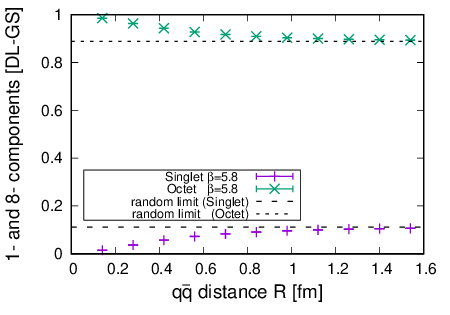}
\caption{\label{SO_DL1}
Singlet and octet components ($\rho_{\bm s, \bm s}$ and $\rho_{\bm a, \bm a}$) for $\Gamma=\CHDLa$ channel plotted as a function of the interquark distance $R$. The dashed and dotted lines indicate the values 1/9 and 8/9 for $\rho_{\bm s, \bm s}$ and $\rho_{\bm a, \bm a}$ in the random-limit color configuration, respectively.}
\end{figure}

\begin{figure}[h]
 \includegraphics[width=8cm]{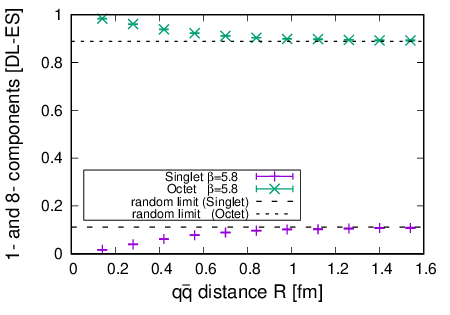}
\caption{\label{SO_DL2}
Singlet and octet components ($\rho_{\bm s, \bm s}$ and $\rho_{\bm a, \bm a}$) for $\Gamma=\CHDLb$ channel plotted as a function of the interquark distance $R$. The dashed and dotted lines indicate the values 1/9 and 8/9 for $\rho_{\bm s, \bm s}$ and $\rho_{\bm a, \bm a}$ in the random-limit color configuration, respectively.}
\end{figure}

Figure~\ref{SO_SG1} shows the $R$ dependence of 
the singlet and octet components $\rho_{\bm s, \bm s}$ and $\rho_{\bm a, \bm a}$
for the $\CHSGa$ channel,
and Figs.~\ref{SO_SG2}-\ref{SO_DL2}
demonstrate those for the excited channels, $\CHSGb$, $\CHPIa$, $\CHPIb$, $\CHDLa$ and $\CHDLb$, respectively.
One can find that
a $q\bar q$ pair in the $\CHSGa$ channel
forms a purely {\it color singlet} configuration at $R\rightarrow 0$,
and the ratio $\rho_{\bm s,\bm s}:\rho_{\bm a,\bm a}$ 
approaches $1:8$ of the random color configuration as $R$ increases.
The $R$ dependence is consistent with the previous work~\cite{Takahashi:2019ghj},
and we confirm that the diagonalization process of the Hamiltonian 
is successful.
This result indicates that in-between flux-tube formation is 
expressed by the color leak from color sources to a flux tube,
which can be quantified as the color screening effect among color sources, quarks.
It is of great interest that $q\bar q$ pairs 
in all the gluonically excited channels
form a purely {\it color octet} configuration at $R\rightarrow 0$.
It is consistent with a simple gluonic excitation picture,
where a $q\bar q$ pair and one constituent gluon are bound.
The random color configuration is mixed as $R$ increases,
and in the limit of  $R\rightarrow \infty$, 
the ratio of $\rho_{\bm s,\bm s}$ and $\rho_{\bm a_i,\bm a_i}$ 
again approaches that for the random color configuration, 
$\rho_{\bm s,\bm s}:\rho_{\bm a,\bm a}=1:8$.
This observation supports the scenario
that the color screening effect by the in-between flux tube occurs
also in the excited $q\bar q$ systems.

As shown in the results of $\rho_{\bm s,\bm s}$ and $\rho_{\bm a,\bm a}$ for $\CHSGa$ channel, the speed at which the color configuration 
approaches the random limit in this channel
is the slowest in all the channels,
meaning that the singlet $q\bar q$ correlation remains to some extent.
More quantitative discussion of the quenching speed is given in the next subsection.

For the lowest excitation channel, $\CHPIa$, the color components at small $R$ regions
are dominated by the octet components,
which means that a $q\bar q$ pair forms a color octet configuration at $R\rightarrow 0$ (See Fig.~\ref{SO_PI1}).
It is consistent with the constituent gluon picture for gluonic excitation states,
where a color octet $q\bar q$ pair and a color octet constituent gluon 
are bound to form a totally color singlet physical state.
At the large $R$ regions, 
a random color configuration is mixed like $\CHSGa$ channel and the ratio of singlet and octet components again approaches $1:8$.
It means that even in the channels with gluonic excitations,
the flux-tube formation emerges as the mixture of random color configurations.

The color components for
$\CHSGb$, $\CHPIb$, $\CHDLa$, $\CHDLb$ channels shown
in Figs.~\ref{SO_SG2},\ref{SO_PI2},\ref{SO_DL1},\ref{SO_DL2}
indicate that these states are also considered to be gluonic excitation states,
since the color configurations of them are dominated by a color octet configuration at small $R$.
Similarly to the case of the $\CHPIa$ channel, a random color configuration
is gradually mixed as $R$ increases, and 
the ratio of singlet and octet components finally approaches $1:8$.
The quenching speed of the color correlation 
in the $\CHSGb$, $\CHPIb$, $\CHDLa$, $\CHDLb$ channels is faster
compared with that in the $\CHSGa$ and $\CHPIa$ channels.
A closer look reveals a tiny $R$-independent contribution in the $\CHSGb$ channel:
the ratio of singlet and octet components slightly deviates from $1:8$ at the large $R$ regions,
which is clarified in detail in the later sections.


\subsection{$R$ dependence of $F(R)$}

It is naively expected that a $q\bar q$ pair in $\CHSGa$ (ground state) channel
forms a color singlet configuration at $R\rightarrow 0$~\cite{Takahashi:2019ghj,Takahashi:2020bje},
whereas those for other channels
($\CHSGb$, $\CHPIa$, $\CHPIb$, $\CHDLa$, $\CHDLb$)
are expected to
form an octet color configuration at $R\rightarrow 0$,
since excited gluon fields carry adjoint colors.
These behaviors can be confirmed in Figs.~\ref{SO_SG1}-\ref{SO_DL2}.
Then we expect that
the ansatz $\rho_{0,{\bm s}}^{\rm ansatz}(R)$ works for $\CHSGa$ channel,
and $\rho_{0,{\bm a}}^{\rm ansatz}(R)$ is reasonable for $\CHSGb$, $\CHPIa$, $\CHPIb$, $\CHDLa$, and $\CHDLb$ channels
(See Eqs.~(\ref{ansatzs01})$-$(\ref{ansatza03}))
.

We extract $\FRs$ and $\FRa$ for each channel from the components, $\rho_{\bm s,\bm s}(R)$ and $\rho_{\bm a,\bm a}(R)$, evaluated with lattice QCD calculations.
For example, 
for $\Gamma=\CHSGb$, $\CHPIa$, $\CHPIb$, $\CHDLa$, and $\CHDLb$,
we can compute $\FRa$ as
\begin{equation}
\FRa=1 - N_c^2 \rho(R)_{\bm s,\bm s}.
\end{equation}
Due to the normalization condition
\begin{equation}
 \rho(R)_{{\bm s},{\bm s}}
+
 \rho(R)_{{\bm a},{\bm a}}
=1,
\end{equation}
the independent quantity at a given $R$ is only $\rho_{{\bm s}, {\bm s}}$ or $\rho_{{\bm a}, {\bm a}}$.

\begin{figure}[h]
 \includegraphics[width=8cm]{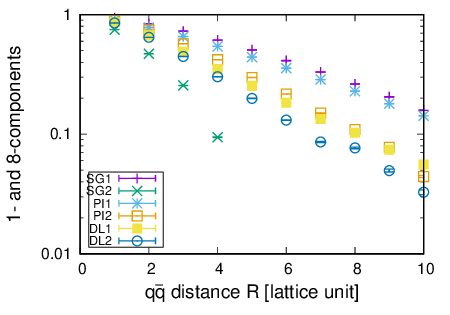}
 \caption{\label{F_ALL}
The logarithmic plot of the fractions of the initial states, $F(R;\Gamma)$ for $\Gamma=\CHSGa[{\rm SG1}]$, $\CHSGb[{\rm SG2}], \CHPIa[{\rm PI1}], \CHPIb[{\rm PI2}], \CHDLa[{\rm DL1}], \CHDLb[{\rm DL2}]$.}
\end{figure}
\begin{figure}[h]
 \includegraphics[width=8cm]{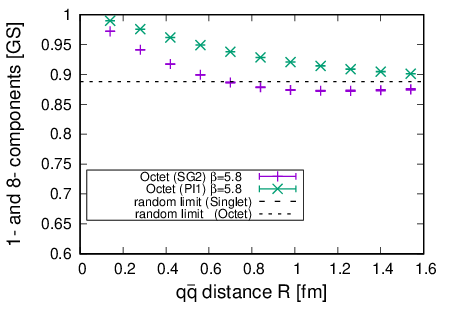}
 \caption{\label{comparison}
A comparison of the octet components $\rho_{\bm a, \bm a}(R;\CHSGb)[{\rm SG2}]$ and $\rho_{\bm a, \bm a}(R;\CHPIa)[{\rm PI1}]$. A negative constant contribution in $\rho_{\bm a, \bm a}(R;\CHSGb)$ can be found in the plot.}
\end{figure}
\begin{figure}[h]
 \includegraphics[width=8cm]{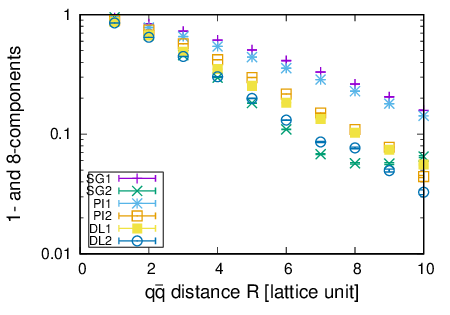}
 \caption{\label{F2_ALL}
The logarithmic plot of the fractions of the initial states, $\tilde F(R;\CHSGb)[{\rm SG2}]$ and $F(R;\Gamma)$ for $\Gamma=\CHSGa[{\rm SG1}]$, $\CHPIa[{\rm PI1}], \CHPIb[{\rm PI2}], \CHDLa[{\rm DL1}], \CHDLb[{\rm DL2}]$.
The ratio $\tilde F(R;\CHSGb)$ is defined as $\tilde F(R;\CHSGb)\equiv F(R;\CHSGb) + \delta$ with the original function $F(R;\CHSGb)$.
}
\end{figure}

We define the fraction of the initial state remaining at $R$ for $\Gamma$ channel as
\begin{eqnarray}
F(R;\Gamma) 
&\equiv&
\FRs
\end{eqnarray}
for $\Gamma=\CHSGa$, and as
\begin{eqnarray}
F(R;\Gamma) 
&\equiv&
\FRa
\end{eqnarray}
for other channels, $\Gamma=\CHSGb$, $\CHPIa$, $\CHPIb$, $\CHDLa$, and $\CHDLb$.
In the case of $\Gamma=\CHSGa$, the initial ($R\rightarrow 0$) color configuration 
is dominated by a color singlet one, as $\lim_{R\rightarrow 0}F_s(R)=1$, 
hence we adopt $\FRs$ instead of $\FRa$.
Thus defined $F(R;\Gamma)$ indicates the component of the color correlation;
it starts from 1 at $R\rightarrow 0$ and decreases as $R$ enlarges.
$F(R;\Gamma)$ calculated for all the channels are logarithmically plotted as a function of the interquark
distance $R$ in Fig.~\ref{F_ALL}. 
Most of them show a linear dependence on $R$, 
which means that initial color correlations are exponentially quenched
as $R$ increases,
and their color configurations approach the random state:
$F(R;\Gamma)$ can be expressed as $F(R;\Gamma)\sim A(\Gamma)\exp(-B(\Gamma)R)$.
However, 
the log plot of $F(R;\CHSGb)$ does not show a clear linear dependence
at all $R$ regions,
and it takes a smaller value than others at small $R$ regions.
This tendency peculiar to $\CHSGb$ channel implies
a negative constant contribution mixed to $\rho_{\bm a,\bm a}$.
Octet components $\rho_{\bm a, \bm a}$ for $\Gamma=\CHSGb$ and $\CHPIa$ ($\rho_{\bm a, \bm a}(R;\CHSGb)$ and $\rho_{\bm a, \bm a}(R;\CHPIa)$)
are compared in Fig.~\ref{comparison}.
$\rho_{\bm a, \bm a}(R;\CHSGb)$ clearly starts from a smaller value than
$\rho_{\bm a, \bm a}(R;\CHPIa)$ at small $R$ regions,
and $\rho_{\bm a, \bm a}(R;\CHSGb)$ seems to approach 
some constant value slightly smaller than $8/9$, which implies that 
a negative $R$-independent contribution exists in $\rho_{\bm a, \bm a}(R;\CHSGb)$:
$F(R;\CHSGb)$ is expressed as $F(R;\CHSGb)\sim A(\CHSGb)\exp(-B(\CHSGb)R)-\delta$.
A numerical fit gives a offset value $\delta=0.2043$, and $\tilde F(R;\CHSGb)\equiv F(R;\CHSGb)+\delta$ plotted along with other $F(R;\Gamma)$'s
is shown in Fig.~\ref{F2_ALL}.
After this correction, the log plot of $\tilde F(R;\CHSGb)$ shows a clear linear $R$ dependence
at $R<1.2$ fm.
On the other hand, $\tilde F(R;\CHSGb)$ again goes up at $R>1.2$ fm .
Taking into account that color configurations evaluated presently are strongly affected
by a finite volume effect, it might be a finite-volume artifact.
In fact, in Fig.11 in Ref.~\cite{Takahashi:2019ghj}, 
one can find considerable finite-volume effect for $R>1.0$ fm.
Such artifacts can be also found in the fitted parameters (Fig.7 in Ref.~\cite{Takahashi:2019ghj}).
The detailed clarification of color correlation at $R\rightarrow \infty$ region
with a huge lattice volume is left for future studies.

The exponential decay of the $q\bar q$ correlation $F(R;\Gamma)$ indicates
the exponential color screening effects due to in-between gluons.
One can find roughly three different magnitudes of slope in Fig.~\ref{F2_ALL}.
The correlation quenching speed for $\CHSGa$ and $\CHPIa$ channels
is the slowest and the color correlation remains at larger $R$.
The quenching speed for $\CHPIb$ and $\CHDLa$ channels is faster
than that for $\CHSGa$ and $\CHPIa$ channels.
Further fast quenching of the color correlation is found in $\CHSGb$ and $\CHDLb$ channels.
In the ground state channel $\CHSGa$,
the color leak from quarks to a flux tube is most suppressed.
The fact that the color screening speed for the lowest excited channel
$\CHPIa$ is comparable with $\CHSGa$ channel in magnitude may indicate that
the $\CHPIa$ state has the simplest gluonic excitation mode
that does not accelerate the color screening effect.
Other excited states are considered 
to have more complicated gluonic excitation
and the color correlation between quarks are easily randomized.

We fit $F(R;\Gamma)$ with an exponential function as
\begin{equation}
F(R;\Gamma)  = A(\Gamma) \exp(-B(\Gamma) R),
\end{equation}
and extract the ``screening mass'' $B(\Gamma)$.
For the $\CHSGb$ channel, the values of $\tilde F(R;\CHSGb)$ after the correction are used for fitting.
The fit range is set to $4\leq R \leq 7$ in lattice unit, 
in which the data show an exponential damping,
as seen in Fig.~\ref{F2_ALL}.
\begin{figure}[h]
\includegraphics[width=08cm]{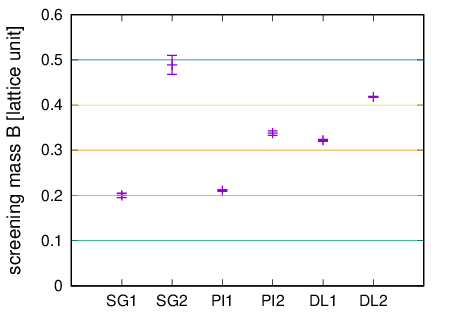}
 \caption{\label{paramB}
The screening masses $B(\Gamma)$ that quantify the color screening effect between static quarks
are plotted for all the channels.
Each $B$ is extracted with an exponential fit
$F(R;\Gamma)  = A(\Gamma) \exp(-B(\Gamma) R)$.
 }
\end{figure}

In Fig.~\ref{paramB}, the fitted parameters $B(\Gamma)$ are plotted,
and they are also listed in Table.~\ref{paramBtable}.
\begin{table}
 \begin{tabular}{c|c|c|c|c|c}
\hline
$\CHSGa$ & $\CHSGb$ & $\CHPIa$ & $\CHPIb$ & $\CHDLa$ & $\CHDLb$ \\
\hline
0.200(5) & 0.489(21)& 0.211(2) & 0.338(5) & 0.322(2) & 0.418(1) \\
\hline
 \end{tabular}
\caption{
\label{paramBtable}
The screening masses $B(\Gamma)$ obtained by an exponential fit,
$F(R;\Gamma)  = A(\Gamma) \exp(-B(\Gamma) R)$,
are listed in lattice unit.
}
\end{table}
One can categorize the screening masses as,
\begin{eqnarray}
B(\CHSGa)\sim B(\CHPIa)
<
B(\CHPIb)\sim B(\CHDLa)
\nonumber \\
<
B(\CHDLb)
<
B(\CHSGb).
\end{eqnarray}
The screening mass for $\CHPIa$ channel is as small as that for $\CHSGa$,
whereas those for other channels are significantly larger.
It might imply that 
in the viewpoint of color screening effects,
the gluonic excitations except for $\CHPIa$ consist of a sum of ``fundamental excitations'' of the $\CHPIa$ channel.

\subsection{Comparison with results at $T>0$}

\begin{figure}[h]
\includegraphics[width=08cm]{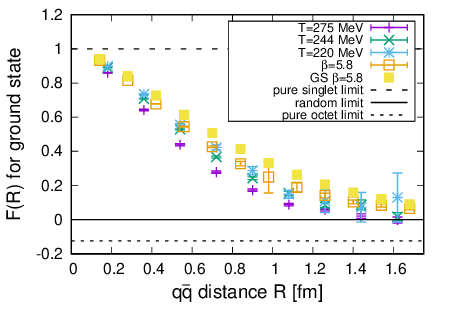}
 \caption{\label{FT}
$F(R)$ at $T>0$
reconstructed from the $F(R;\Gamma)$'s for 
low-lying 6 channels ($\CHSGa$, $\CHSGb$, $\CHPIa$, $\CHPIb$, $\CHDLa$, $\CHDLb$)
and the lattice QCD results shown in Ref.~\cite{Takahashi:2020bje} are plotted.
 }
\end{figure}

In Ref.~\cite{Takahashi:2020bje}, the color structures of a $q\bar q$ pair 
at finite temperature were investigated in detail
by analyzing the color density matrix $\rho$ as well as the corresponding entanglement entropy,
and the $q\bar q$ color correlations were
found to be quickly quenched 
across the deconfinement phase transition temperature $T_c$.
It means the $q\bar q$ color configuration is strongly randomized above $T_c$
as the in-between flux tube disappears.
Even below $T_c$, it was found that 
the color correlation gradually weaken as the temperature rises
due to the temperature effect.
We here clarify if the temperature effect at $T<T_c$ can be explained
as a thermal ensemble based on the present results for excited $q\bar q$ states at $T=0$.
We reconstruct the $F(R)$ at $T>0$
from the $F(R;\Gamma)$'s for 
low-lying 6 channels ($\CHSGa$, $\CHSGb$, $\CHPIa$, $\CHPIb$, $\CHDLa$, $\CHDLb$),
and compare it with the lattice QCD results shown in Ref.~\cite{Takahashi:2020bje}.
We define the density operator $\hat \rho_T$ at finite temperature as
\begin{eqnarray}
\hat \rho_T(R)
=
\sum_{\Gamma}^{\rm 6\  channels}
\hat \rho(R;\Gamma)
e^{-\frac1T (E(R;\Gamma)-E_0(R))},
\label{FTform}
\end{eqnarray}
where $T$ is the temperature of the system
and $E_0(R)$ is a energy of a ground-state $q\bar q$ pair ($E_0(R)=E(R;\CHSGa)$).
In Fig.~\ref{FT}, we show the $F(R)$ reconstructed using Eq.(\ref{FTform})
as well as the $F(R)$'s shown in Ref.~\cite{Takahashi:2020bje}.
Here, the temperature $T$ in Eq.(\ref{FTform}) is chosen to be 250 MeV.
$F(R)$ reconstructed using Eq.(\ref{FTform}) is shown by open squares,
and filled squares denote $F(R)$ at $T=0$.
Other symbols represent $F(R)$'s at each temperature demonstrated in Ref.~\cite{Takahashi:2020bje}.
As is seen in Fig.~\ref{FT}, the lattice QCD data at $T>0$
are reproduced by Eq.(\ref{FTform}) more or less in the range of $0<R<0.8$ fm.
Though the coincidence is lost at larger $R$, 
this deviation is considered to arise from the finite volume effects.
The lattice adopted in Ref.~\cite{Takahashi:2020bje} is smaller ($V=24^3$)
and the finite volume artifact would remain for large $R$.

\section{Summary and concluding remarks}
\label{Sec.Summary}

  We have investigated the color correlation inside
  a static quark and antiquark pair accompanied by gluonic excitations (hybrid $q\bar q$ system)
  in the confined phase by means of lattice QCD.
  We have performed quenched lattice QCD calculations with the Coulomb gauge
  adopting the standard Wilson gauge action,
  and the spatial volume considered here is $L^3 = 32^3$ at $\beta = 5.8$,
  which corresponds to the lattice spacing $a=0.14$ fm 
  and the system volume $L^3=4.5^3$ fm$^3$.
  The reduced density matrix $\rho$ in the color space for a $q\bar q$ system with the interquark distance $R$
  has been constructed from link variables and $\rho$ has been analyzed based on the ansatz
  we proposed in our previous paper.
  The color density matrix $\rho$ of static $q\bar q$ pairs have been computed
  in 6 channels ($\CHSGa$, $\CHSGb$, $\CHPIa$, $\CHPIb$, $\CHDLa$, $\CHDLb$ channels),
  and we have investigated the $R$ dependence of color correlations.

  For the $\CHSGa$ channel, the ground state channel of a static $q\bar q$ system,
  we have confirmed that when $R$ is small a $q\bar q$ pair forms
  a purely {\it color singlet} configuration, which is consistent with our previous study.
  In the case of hybrid $q\bar q$ systems with gluonic excitations
  ($\CHSGb$, $\CHPIa$, $\CHPIb$, $\CHDLa$, $\CHDLb$ channels),
  a $q\bar q$ pair at $R\rightarrow 0$ always forms a purely {\it color octet} configuration.
  This finding is consistent with the constituent gluon picture,
  where gluonic excitations are expressed by constituent gluons
  that couple to a static $q\bar q$ pair.

  In all the cases investigated here, as $R$ increases, 
  an uncorrelated state represented by random color configurations,
  where all the $N_c^2$ components mix with equal weights,
  enters in $\rho$ due to the color-correlation quenching by in-between gluons,
  and the ratio of singlet and octet components in $\rho$ finally approaches $1:8$.

  In order to clarify the quenching speed in a quantitative way,
  we have defined and evaluated the "screening mass"
  from the slope of exponential damping of the color correlation.
  The screening masses $B$ for 6 channels satisfy
  \begin{eqnarray*}
  B(\CHSGa)\sim B(\CHPIa)
  <
  B(\CHPIb)\sim B(\CHDLa)
  <
  B(\CHDLb)
  <
  B(\CHSGb),
  \end{eqnarray*}
  The screening mass for $\CHPIa$ channel is as small as that for $\CHSGa$,
  whereas those for other channels are significantly larger.
  It might imply that the gluonic excitations except for $\CHPIa$ consist of 
  a sum of ``fundamental excitations'' of the $\CHPIa$ channel in the viewpoint of color screening effects.
  In the ground-state channel $\CHSGa$,
  the color leak from quarks to a flux tube is most suppressed,
  and the fact that the color screening speed for the lowest excited channel
  $\CHPIa$ is the same in magnitude as $\CHSGa$ channel may indicate that
  the $\CHPIa$ state has the simplest gluonic excitation mode
  that does not accelerate the color screening effect.
  Other excited states are considered to have more complicated gluonic excitation
  and the color correlation between quarks are easily randomized as $R$ increases.
  {It would be interesting to clarify the possible relationship 
  between the screening mass and the gluelump mass~\cite{Foster:1998wu,Marsh:2013xsa,Berwein:2015vca,Herr:2023xwg}}.

  {We have also tried to reproduce the color density matrix of a $q\bar q$ pair at finite temperature 
  using the density matirices for low-lying 6 channels at $T=0$ 
  ($\CHSGa$, $\CHSGb$, $\CHPIa$, $\CHPIb$, $\CHDLa$, $\CHDLb$ channels),
  and have found that the lattice QCD data at $T>0$ are reproduced by the thermal average in the range of $0<R<0.8$ fm.
  Note that the color structure of a $q\bar q$ system in hot medium has been also discussed in different literatures~\cite{Brambilla:2017zei,Bazavov:2018wmo,Brambilla:2022ynh}.
  }

  A future direction for this work would be the analysis of multiquark systems.
  Multiquark systems are attracting much attention,
  and clarification of their internal structures has been of great importance.
  Our method can be easily extended to the analysis of multiquark systems,
  which is now in progress.

\begin{acknowledgements}
This work was partly supported by
Grants-in-Aid of the Japan Society for the Promotion of
Science (Grant Nos. 18H05407, 22K03608, 22K03633).
\end{acknowledgements}

\end{document}